\documentclass[journal=jacsat,manuscript=article]{achemso}

\usepackage[version=3]{mhchem}

\usepackage{placeins}
\usepackage{booktabs}
\usepackage{multirow}
\usepackage{arydshln}
\usepackage{threeparttable}
\usepackage{hyperref}

\author{Jens Wagner}
\author{Zeno Romero}
\author{Kerstin M\"unnemann}
\author{Sebastian Schmitt}
\author{\newline Thomas Specht}
\author{Hans Hasse}
\author{Fabian Jirasek}
\email{fabian.jirasek@rptu.de}
\phone{+49 (0)631 - 205 4685}
\affiliation[Unknown University]
{Laboratory of Engineering Thermodynamics (LTD), RPTU Kaiserslautern, Germany}

\title[An \textsf{achemso} demo]
  {Hybrid Machine Learning for Enhanced Prediction of Diffusion Coefficients in Liquids}

\begin{document}

\begin{abstract}
Diffusion coefficients are key thermophysical properties for modeling mass transport in liquids, but experimental data are scarce, making reliable prediction methods indispensable.
In the present work, we introduce a new method for predicting diffusion coefficients of molecular components at infinite dilution in pure liquid solvents by integrating the Stokes–Einstein~(SE) equation with machine learning~(ML).
Unlike previous ML approaches, the resulting hybrid Enhanced Stokes-Einstein~(ESE) model provides strictly physically consistent predictions for diffusion coefficients as a function of temperature across a broad range of binary mixtures.
Trained and validated using an extensive compilation of literature data for infinite-dilution diffusion coefficients in binary liquid systems, ESE achieves significantly higher prediction accuracies than the previous state-of-the-art model, SEGWE, while requiring only the SMILES strings encoding of the molecular formulae of the components of interest as additional inputs, which are always available.
This simplicity makes ESE broadly applicable, e.g., for process design and optimization.
The ESE model and its source code are fully disclosed and are directly accessible via an interactive web interface at \href{https://ml-prop.mv.rptu.de/}{\mbox{https://ml-prop.mv.rptu.de/}}.
\end{abstract}

\section{Introduction}

Diffusion is a ubiquitous phenomenon that governs mass transport and plays a central role in a wide range of natural and technological processes. Accurate diffusion coefficients are therefore essential for the quantitative description and modeling of transport phenomena. However, determining diffusion coefficients experimentally is often demanding and time-consuming, and available data remain sparse for many relevant systems. As a consequence, reliable, broadly applicable predictive models of diffusion coefficients are indispensable. In this context, diffusion in liquids is particularly relevant, as it is both challenging to predict and critically important for many applications, notably reaction and separation processes in chemical engineering.

In general, two types of diffusion must be distinguished: \textit{Mutual diffusion} refers to the motion of \textit{collectives} of molecules of different components in mixtures and can be described by the Fickian or the Maxwell–Stefan approach, each associated with its own set of diffusion coefficients. In contrast, \textit{self-diffusion} relates to the Brownian motion of \textit{individual} molecules, which is defined in pure components and mixtures. In the limit of infinite dilution of a studied solute in a specific solvent, the differences between the Fickian and the Maxwell–Stefan mutual diffusion coefficients and the self-diffusion coefficient of the solute vanish. Consequently, the term 'diffusion coefficient at infinite dilution' collectively describes both mutual and self-diffusion at infinite dilution and is used this way in the present work. 

Data on diffusion coefficients at infinite dilution are directly relevant for modeling mixtures in which the diffusing component is highly diluted. Furthermore, the data at infinite dilution can be used to predict Maxwell–Stefan mutual diffusion coefficients at finite concentrations in binary and multicomponent mixtures, e.g., using the method of Vignes~\cite{Vignes1966, Kooijman1991}. However, even for binary mixtures, i.e., a single solute in a pure solvent, experimental data on diffusion coefficients at infinite dilution are remarkably scarce~\cite{Gromann2022}, making the accurate prediction of these coefficients essential.

Diffusion coefficients~$D^{\infty}_{ij}$ of a solute~$i$ infinitely diluted in a liquid solvent~$j$ can, under certain assumptions, be described by the Stokes-Einstein~(SE) equation~\cite{Einstein1905}, which is derived considering the motion of a hard sphere~(diffusing particle) in a continuous fluid~(solvent):

\begin{equation}
D^{\infty, \mathrm{SE}}_{ij} = \frac{k_B T}{6 \pi \eta_j r_i}
\label{eq_SEE}
\end{equation}

where $k_B$ is the Boltzmann constant, $T$ is the absolute temperature, $\eta_j$ is the dynamic viscosity of the solvent~$j$ at the temperature $T$, and $r_i$ is the effective radius of the solute~$i$. The latter can be estimated from the molar mass of the solute~$M_i$, its specific density~$\rho_i$, and the Avogadro constant~$N_A$, leading to:

\begin{equation}
r_i = \sqrt[3]{\frac{3 M_i}{4 \pi \rho_i N_\mathrm{A}}f}
\label{eq_ri}
\end{equation}

where $f$ is an empirical correction factor of the spheric volume (sometimes called packing fraction) that is often found to be in the range of $f = 0.6\,\,\ldots\,\,0.8$. For predictive applications in liquids, often a default value of $f = 0.64$ is used~\cite{Evans2013}.

Despite the simplicity of the underlying model, the SE equation captures many essential features of diffusion coefficients. However, it exhibits significant deficiencies in quantitatively predicting diffusion coefficients in real liquid mixtures. To address these deficiencies, many empirical extensions of the SE model have been proposed in the literature~\cite{Taylor1993, Wilke1955, Reddy1967, Tyn1975, Evans2013}. In a recent comparison of the predictive performance of these extensions, the Stokes-Einstein~Gierer-Wirtz~Estimation~(SEGWE) of Evans et al.~\cite{Evans2013, Evans2018} was found to be the most accurate of the semi-empirical literature models~\cite{Gromann2022}. 

In the SEGWE model, the extensions to the original SE model account for the relative size of the solvent compared to the solute by incorporating the solvent's molar mass and introducing an empirical parameter, the so-called effective density, $\rho_{\mathrm{eff}} = 627$~kg~m$^{-3}$, which was fitted to experimental data for a variety of binary mixtures. Although these modifications significantly improve the predictive performance of the SEGWE model relative to the SE model across many systems, relying on a single global empirical parameter is insufficient to capture the range of possible interactions in mixtures, as already noted by the developers of SEGWE~\cite{Evans2018}. These results are consistent with findings from our recent works~\cite{Specht2021, Specht2023a, Specht2023b, Specht2023c, Wagner2025, Wagner2026}, which indicate that the SEGWE model exhibits several deficiencies, including an inadequate treatment of polar interactions.

As an alternative to the available semi-empirical models, data-driven quantitative structure-property relationships (QSPR) and machine learning (ML) approaches for the prediction of diffusion coefficients at infinite dilution~\cite{Khajeh2011, Abbasi2014, Mariani2020, Aniceto2021, Aniceto2024} and finite concentrations~\cite{Beigzadeh2012, Abbasi2014} in binary mixtures have been proposed. While these studies successfully identified solute and solvent properties that influence diffusion, which were then used as inputs, they often achieved this only by limiting their analysis to specific solvent classes. For example, Aniceto et al.~\cite{Aniceto2021} developed separate models for polar and nonpolar solvents, although categorizing components into these classes is often ambiguous. Other authors focused exclusively on water~\cite{Khajeh2011, Aniceto2024} or hydrocarbons~\cite{Abbasi2014} as solvents. Furthermore, these fully data-driven approaches lack built-in physical constraints, e.g., ensuring that the diffusion coefficient increases with temperature, which may lead to unphysical results, especially when applied outside the conditions of the experimental data used for model training.

Matrix completion methods (MCMs) constitute a special class of ML methods for predicting properties of binary mixtures that do not rely on molecular descriptors. MCMs exploit the fact that the properties of binary mixtures can conveniently be stored in matrices, where rows and columns correspond to the components of the mixture, thereby formulating the prediction of the properties of unstudied mixtures as a matrix completion problem. Purely data-driven MCMs, which only learn from the available experimental mixture data, and hybrid MCMs, which incorporate physical knowledge and physical models, have been developed for the prediction of various mixture properties~\cite{Jirasek2020, Jirasek2020b, Damay2021, Jirasek2022, Hayer2022, Jirasek2023b, Jirasek2023c, Hayer2024, Hoffmann2024, Gond2025, Hayer2025cosmo, Zenn2025, Hayer2025}, including diffusion coefficients at infinite dilution in binary mixtures~$D^{\infty}_{ij}$~\cite{Gromann2022, Romero2025}. 

Hybrid models combining data-driven MCMs with the SEGWE model showed improved prediction accuracy for $D^{\infty}_{ij}$ compared to all available semi-empirical models~\cite{Gromann2022}, including the SEGWE model alone. However, the downside of MCMs, including the hybrid one based on the SEGWE model, is that they are restricted to mixtures of solutes and solvents for which at least some experimental $D^{\infty}_{ij}$, in combination with other solvents or solutes, are available. Furthermore, MCMs, in their basic form, are restricted to a single temperature.

The second limitation can be addressed by extending MCMs to tensor completion methods~(TCMs)~\cite{Romero2026}, which capture temperature dependence by treating temperature as an additional tensor dimension. However, these TCMs are still restricted to solutes and solvents for which at least some experimental data are available. Thus, they cannot be used to predict~$D^{\infty}_{ij}$ for systems containing solutes or solvents without corresponding experimental training data. 

To summarize, models that allow for the physically consistent and accurate prediction of liquid-phase diffusion coefficients at infinite dilution, $D^{\infty}_{ij}$, across temperatures and across a broad range of solutes and solvents, including those that are completely unstudied, are currently missing. 

In this work, we introduce such a method by integrating the physical Stokes-Einstein equation with a neural network~(NN), resulting in our hybrid Enhanced Stokes-Einstein~(ESE) model. Trained on experimental $D^{\infty}_{ij}$ data, the NN generates a mixture-specific scaling factor by leveraging simple molecular descriptors of both the solute and solvent, which can be readily and automatically derived from the SMILES~\cite{Weininger1988} strings of the components. Crucially, the NN is constrained to preserve the physical principles embedded in the SE model, thereby ensuring that the hybrid ESE model yields physically consistent predictions across all temperatures while leveraging the flexibility and predictive power of the ML algorithm. We benchmark ESE against the SEGWE, an MCM~\cite{Romero2025}, and a TCM~\cite{Romero2026} model, demonstrating its superior predictive performance across a wide range of mixtures and temperatures.

\section{Methods}

\subsection{Model Architecture}

\begin{figure}
\centering
    \includegraphics[width=13cm]{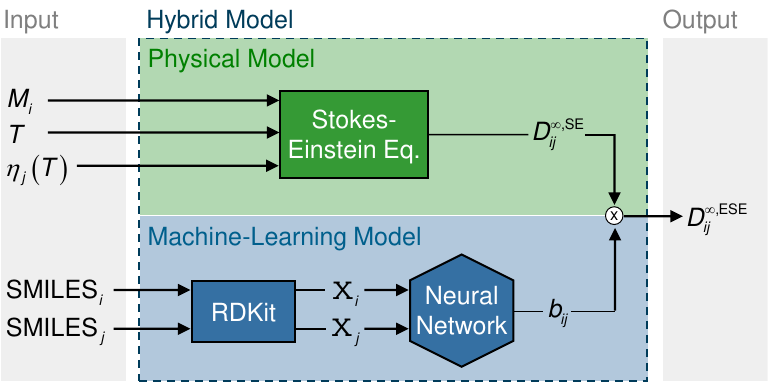}
    \caption{Schematic overview of the hybrid Enhanced Stokes-Einstein~(ESE) model for predicting liquid-phase diffusion coefficients at infinite dilution in binary mixtures~$D^{\infty}_{ij}$. The prediction of the Stokes-Einstein~(SE) equation,~$D^{\infty,\mathrm{SE}}_{ij}$, is thereby corrected using a learned mixture-specific scaling factor, $b_{ij}$, computed by a positively-restricted neural network from the molecular descriptor vectors,~$\mathbf{X}_i$ and $\mathbf{X}_j$~(cf. Tab.~\ref{tab_x}), which are generated for solute~$i$ and solvent~$j$ from their SMILES strings using RDKit~\cite{RDKit2024}.}
    \label{fig_Met}
\end{figure}

Figure~\ref{fig_Met} provides a schematic overview of the hybrid Enhanced Stokes-Einstein~(ESE) model developed in this work, which integrates the Stokes-Einstein equation~(SE) with an NN to improve the prediction of diffusion coefficients at infinite dilution $D^{\infty}_{ij}$ in binary mixtures. In this architecture, first the SE equation~(1) is used to obtain a preliminary prediction $D^{\infty,\mathrm{SE}}_{ij}$ based on the solute molar mass~$M_i$, the temperature~$T$, and the solvent viscosity~$\eta_j$ at the temperature~$T$. To eliminate the need for additional thermophysical property inputs, the solute density was fixed at $\rho_i = 1050\ \mathrm{kg\ m^{-3}}$ and the empirical correction factor at $f = 0.64$ for all solutes~\cite{Evans2013}. Preliminary tests showed that the choice of these values is not critical and has only a minor effect on the results from the trained model, as long as they are within a physically reasonable range. We therefore do not consider them as adjustable hyperparameters here.

\FloatBarrier

Parallel to this, the SMILES strings of the solute~$i$ and solvent~$j$ are processed using the open-source toolkit RDkit~\cite{RDKit2024} to automatically generate their molecular descriptor vectors, $\mathbf{X}_i$ and $\mathbf{X}_j$, respectively, which are described in detail below~(cf. Tab.~\ref{tab_x}). These molecular descriptor vectors serve as sole inputs to an NN, which outputs the mixture-specific scaling factor~$b_{ij}$, which is independent of temperature. In combination with the NN being restricted to produce only positive outputs for $b_{ij}$, this ensures that the physical background of the SE equation, specifically regarding the temperature dependence of $D^{\infty}_{ij}$, is retained in the hybrid model. The final prediction, $D^{\infty,\mathrm{ESE}}_{ij}$, is then obtained by multiplying the physical prediction,~$D^{\infty,\mathrm{SE}}_{ij}$, by the respective scaling factor~$b_{ij}$:

\begin{equation}
D^{\infty,\mathrm{ESE}}_{ij} = b_{ij} \cdot D^{\infty,\mathrm{SE}}_{ij}
\label{eq_ESE}
\end{equation}

The NN architecture, as determined from the hyperparameter optimization (see below), comprises two fully connected layers with 32 and 16 nodes, respectively, employs the Rectified Linear Unit~(ReLU) activation function, and utilizes a Softplus activation in the output layer to ensure strictly positive values of $b_{ij}$.

\subsection{Molecular Descriptors}

Table~\ref{tab_x} gives an overview of the molecular descriptors contained in the $\mathbf{X}$ vectors, encoding structural and polarity-related characteristics of the molecules, while allowing unambiguous and automatic derivation from their SMILES strings using RDKit. The same set of descriptors was chosen for the solute and the solvent.

The descriptor set was deliberately kept small and comprises only a limited number of properties to enable effective training on the limited available experimental data for~$D^{\infty}_{ij}$ and to increase the interpretability of the model and its results.
The selection of these descriptors is physically motivated and derived from the molecular properties we identified as having significant influence on diffusion in our previous studies~\cite{Specht2023b, Wagner2025}.
Details on the specific RDKit functions used to obtain these descriptors are provided in Table~S.1 in the Supporting Information.

\begin{table}
\centering
\caption{Molecular descriptors included in the vectors~$\mathbf{X}_i$ and $\mathbf{X}_j$ describing the solute~$i$ and solvent~$j$, respectively, and used as input for the ESE model.}
\label{tab_x}
    \begin{tabular}{ll}
        \toprule
        Label & Molecular descriptor \\
        \midrule
        $M$ & Molar mass in kg~mol$^{-1}$\\
        $R$ & Boolean variable indicating presence of molecular ring structures\\
        $r_\mathrm{Het}$ & Ratio of number of heteroatoms to non-hydrogen atoms\\
        $r_\mathrm{Hal}$ & Ratio of number of halogen atoms to non-hydrogen atoms\\
        $r_\mathrm{Acc}$ & Ratio of number of hydrogen-bond acceptors to non-hydrogen atoms\\
        $r_\mathrm{Don}$ & Ratio of number of hydrogen-bond donors to non-hydrogen atoms\\
        \bottomrule
    \end{tabular}
\end{table}
\FloatBarrier

\begin{figure}
\centering
    \includegraphics[width=12.0cm]{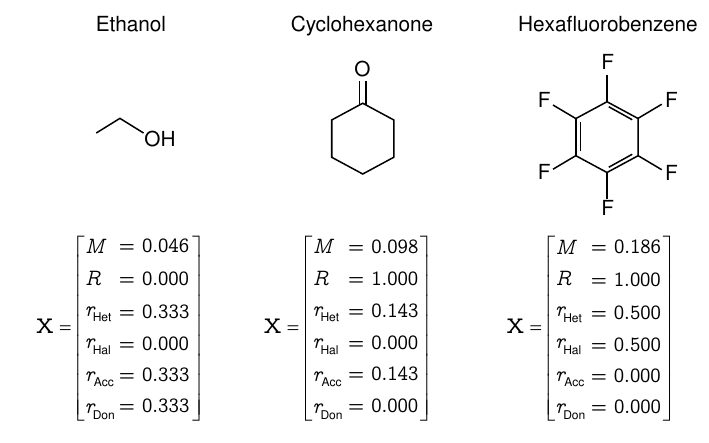}
    \caption{Molecular structures and corresponding molecular descriptor vectors~$\mathbf{X}$ (cf. Table~\ref{tab_x}) exemplary shown for ethanol, cyclohexanone, and hexafluorobenzene.}
    \label{fig_Dec}
\end{figure}

Figure~\ref{fig_Dec} provides the molecular structures along with the corresponding molecular descriptor vectors~$\mathbf{X}$ for three exemplary molecules from our data set.

\FloatBarrier

\subsection{Experimental Database}

In this work, we exclusively consider binary liquid mixtures of non-ionic organic molecules (including water) with molar masses up to 1000~g~mol$^{-1}$, containing no heavier atoms than chlorine.

Experimental data for $D^{\infty, \mathrm{exp}}_{ij}$ were primarily obtained from the database compiled by Großmann et al.~\cite{Gromann2022}, which was subsequently extended to other temperatures by Romero et al.~\cite{Romero2025}. In addition, we supplemented the database with recent results from the literature~\cite{Bellaire2022, Mross2024, Wagner2025, Romero2026}, new available data from the Dortmund Data Bank~(DDB)~\cite{DDB2024}, and some previously unpublished data that were obtained by pulsed field gradient~(PFG) NMR spectroscopy~\cite{Bellaire2020} in our lab. Details on the experimental procedure and the measured data are provided in the Supporting Information.

The total database used in this work contains $N = 1011$ experimental data points in the temperature range from 273.2~K to 363.0~K for 538 binary mixtures comprising 209 unique solutes and 42 unique solvents. The experimental dynamic viscosities of all solvents at the respective temperatures, which are required for the application of all considered models, were obtained from the DDB.

\subsection{Training and Evaluation}

ESE was trained and evaluated using a $K$-fold cross-validation~(CV)~\cite{Stone1974} based on solute-wise data splits, where $K$ corresponds to the number of distinct solutes in our database. For this purpose, in each fold, all data associated with one specific solute were withheld and used as test data, while the remaining data were randomly split data point-wise into training data (80\%) and validation data (20\%). This procedure was repeated for all solutes until predictions for all data points were obtained.

During the training of ESE, the weights of the NN were optimized to predict $b_{ij}$ (cf. Eq.~\ref{eq_ESE}) by minimizing the mean squared relative error~(MSRE), which served as the loss function and is defined as the average over all considered data points of the squared relative error (SRE) between the predicted and experimental values:

\begin{equation} \mathrm{SRE}_{ij} = \left( \frac{D^{\infty,\mathrm{pred}}_{ij} - D^{\infty,\mathrm{exp}}_{ij}}{D^{\infty,\mathrm{exp}}_{ij}} \right)^2 \label{eq_SRE} \end{equation}

For weight updates, the AdamW~\cite{Loshchilov2017} optimizer was employed. After 25 epochs without improvement in the validation loss, the learning rate was reduced by a factor of 0.1. Early stopping was triggered if the validation loss did not improve for 50 consecutive epochs, and the model with the lowest validation loss was selected for final evaluation on test data. 

Hyperparameter optimization was performed using a grid search approach on the validation loss. The optimized hyperparameters, along with a sensitivity analysis and the corresponding validation loss results, are provided in the Supporting Information.

In addition to SRE and MSRE, which were used for model optimization, the predictive performance was also assessed using the absolute relative error (ARE):

\begin{equation} \mathrm{ARE}_{ij} = \left| \frac{D^{\infty,\mathrm{pred}}_{ij} - D^{\infty,\mathrm{exp}}_{ij}}{D^{\infty,\mathrm{exp}}_{ij}} \right| \label{eq_ARE} \end{equation}

and its average over all data points, the mean absolute relative error (MARE), on the test data.

While the above-described data-splitting procedure was used to generate the results shown in the next section to evaluate the predictive capacity of the developed hybrid model, we also provide a 'final' version of ESE. For this final model, we trained an ensemble of ten models~\cite{Dietterich2000} by randomly splitting our dataset, data point-wise, ten times into 95\% training and 5\% validation data and using each split to train ten individual models starting from different initializations, which were subsequently combined into an ensemble by simple averaging of the individual models' predictions. The final ESE model is available on Zenodo~(\href{https://doi.org/10.5281/zenodo.18787099}{\mbox{https://doi.org/10.5281/zenodo.18787099}}) and can also be accessed via our interactive website MLPROP~\cite{Hoffmann2025} (\href{https://ml-prop.mv.rptu.de/}{\mbox{https://ml-prop.mv.rptu.de/}}).

Model training was conducted on an A40 GPU. All models, along with the training and evaluation scripts, were implemented in Python~3.12.7 using PyTorch~2.2.1~\cite{Paszke2019}. Typical training times per fold ranged from 60 to 120~s.

\section{Results and Discussion}

In the following, we discuss the performance of the ESE model for predicting diffusion coefficients at infinite dilution $D^{\infty}_{ij}$ by evaluating it on the test data, i.e., for unseen solutes, and benchmarking its results against the predictions obtained using SEGWE~\cite{Evans2018}. Additionally, the comparison includes predictions from the SE model using the fixed values $\rho_i = 1050\ \mathrm{kg\ m^{-3}}$ and $f = 0.64$, which corresponds to the variant also used within the ESE framework. Further comparisons of ESE to an MCM~\cite{Romero2025} and a TCM for predicting $D^{\infty}_{ij}$, within their restricted application domains, are provided in the Supporting Information.

\begin{figure}
\centering
    \includegraphics[width=8.0cm]{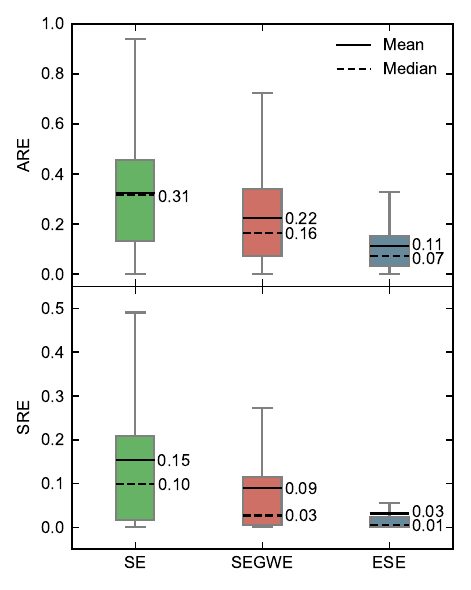}
    \caption{Boxplots of the absolute relative error~(ARE, top) and squared relative error~(SRE, bottom) of the predicted diffusion coefficients at infinite dilution $D^{\infty}_{ij}$ from SE, SEGWE, and ESE. The box width indicates the interquartile range, and the whisker length is 1.5 times the interquartile range. Outliers are not depicted for visual clarity.}
    \label{fig_Box}
\end{figure}

Figure~\ref{fig_Box} shows the ARE (top) and SRE (bottom) calculated by comparing the predicted $D^{\infty}_{ij}$ to the experimental test data in boxplots. The results demonstrate that ESE significantly outperforms the SEGWE model in both error scores. Specifically, ESE halves the mean and median ARE compared to SEGWE and reduces the mean and median SRE by a factor of three relative to the benchmark model, particularly indicating a significantly reduced number of very poorly predicted test data points with ESE. 

\FloatBarrier

\begin{figure}
\centering
    \includegraphics[width=12.5cm]{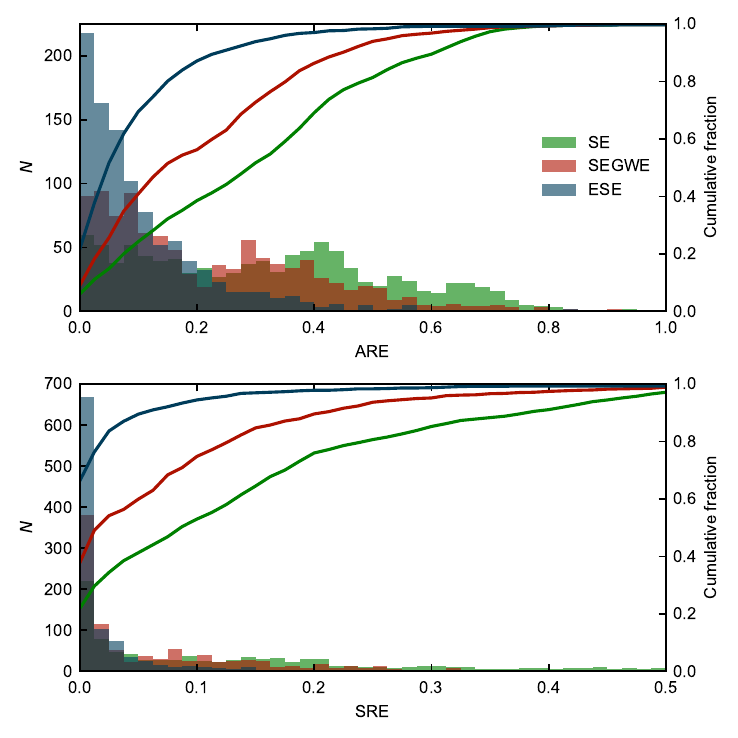}
    \caption{Histograms (bars) and cumulative fractions (lines) showing the number of $D^{\infty}_{ij}$ predicted with a certain absolute relative error (ARE, top) or squared relative error (SRE, bottom) with SE, SEGWE, and ESE. The shown range for the ARE covers $>99~\%$ of the predictions for all models. The shown range of the SRE covers 96.53~\% of the SE predictions, 98.51~\% of the SEGWE predictions, and 99.31~\% of the ESE predictions.}
    \label{fig_Cum}
\end{figure}

Figure~\ref{fig_Cum} compares the accuracy of ESE, SEGWE, and SE for predicting $D^{\infty}_{ij}$ in a histogram indicating the number of test data points that can be predicted with a certain ARE (top) and SRE (bottom). In addition, the cumulative fractions, i.e., the proportions of test data points predicted with error scores smaller than the specified values, are shown. 

\FloatBarrier

The results underpin the improved predictive accuracy of ESE compared to SEGWE and SE. For example, while only approximately 18~\% of the data points can be predicted with an $\mathrm{ARE}< 0.05$ with SEGWE, which is in the range of typical experimental uncertainties reported for $D^{\infty}_{ij}$~\cite{Gromann2022}, ESE can predict approximately 38~\% of the data with this accuracy. 

\begin{figure}
\centering
    \includegraphics[width=16cm]{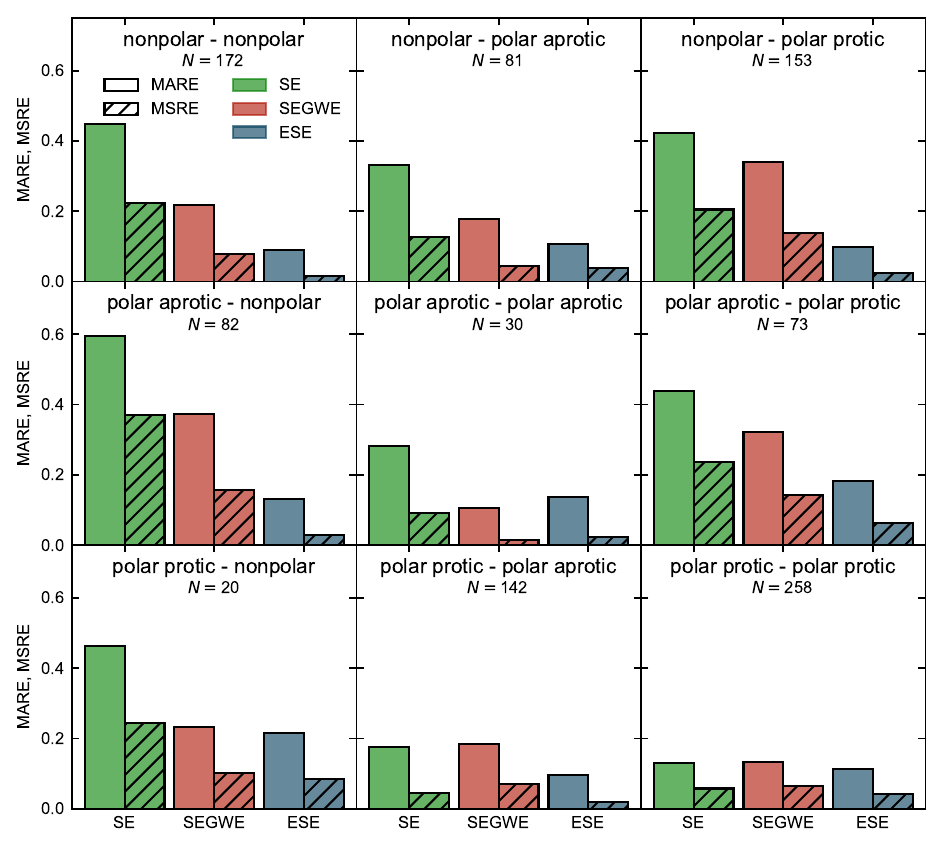}
    \caption{Mean absolute relative error~(MARE) and mean squared relative error~(MSRE) of the predicted $D^{\infty}_{ij}$ from SE, SEGWE, and ESE for nine distinct solute-solvent classes, cf. text. $N$ specifies the number of test data points in our database for each class.}
    \label{fig_Hist}
\end{figure}

For a more detailed analysis of the predictive performance of ESE and the reference models across different types of mixtures, the solutes and solvents were categorized as nonpolar, polar aprotic, or polar protic based on their ratio of hydrogen-bonding acceptor and donor sites, $r_\mathrm{Acc}$ and $r_\mathrm{Don} = 0$, respectively. Components with $r_\mathrm{Acc} = 0$ and $r_\mathrm{Don} = 0$ were thereby classified as nonpolar, components with $r_\mathrm{Acc} \neq 0$ and $r_\mathrm{Don} = 0$ were classified as polar aprotic,
and components with $r_\mathrm{Acc} \neq 0$ and $r_\mathrm{Don} \neq 0$ were classified as polar protic. Figure~\ref{fig_Hist} shows the MARE and MSRE of the predictions for each of the nine resulting mixture classes combining nonpolar, polar aprotic, or polar protic solutes~$i$ and solvents~$j$, along with the corresponding number of data points~$N$ in each class~$i-j$ in our dataset. 

\FloatBarrier

ESE achieves the best performance, i.e., the lowest error scores, across nearly all mixture classes, with the most pronounced advantage over the reference models observed for classes involving nonpolar components.
Only for the polar~aprotic~–~polar~aprotic mixture class, SEGWE yields marginally lower error scores than ESE. Notably, SE provides better predictions than SEGWE for the polar~protic~–~polar~aprotic and polar~protic~–~polar~protic mixture classes.

\begin{figure}
\centering
    \includegraphics[width=15.0cm]{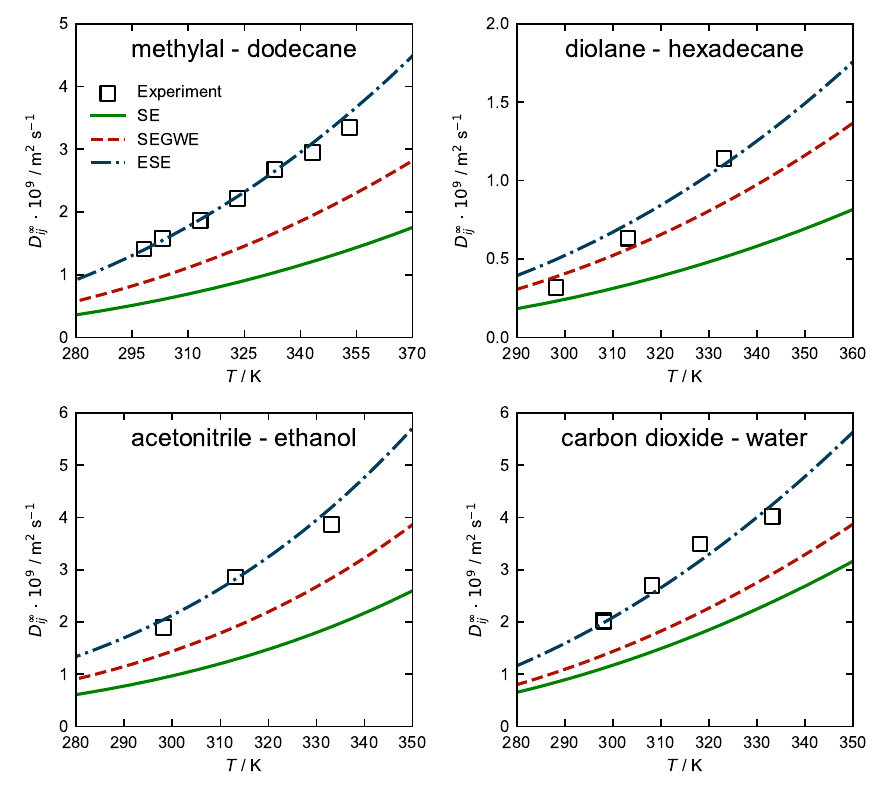}
    \caption{Diffusion coefficients at infinite dilution~$D^{\infty}_{ij}$ as a function of the temperature~$T$ for exemplary mixtures. Symbols represent experimental test data, while lines denote predictions from SE, SEGWE, and ESE.}
    \label{fig_Exp}
\end{figure}

Figure~\ref{fig_Exp} shows $D^{\infty}_{ij}$ predicted with SE, SEGWE, and ESE as a function of the temperature for four exemplary mixtures together with the respective experimental test from our database. To enable these temperature-dependent predictions, viscosity correlations for the respective solvents from the NIST WebBook~\cite{NIST2025} (dodecane~\cite{Lemmon2004} and water~\cite{Huber2009}) or the literature (hexadecane~\cite{Klein2019} and ethanol~\cite{Gonalves2010}) were employed.

For the mixtures methylal~-~dodecane, acetonitrile~-~ethanol, and carbon~dioxide~-~water, the ESE predictions agree very well with the experimental test data, demonstrating exceptional accuracy of ESE in predicting $D^{\infty}_{ij}$ for mixtures with unseen solutes. In contrast, SEGWE consistently underestimates the experimental diffusion coefficients at infinite dilution for these mixtures, whereas SE exhibits an even greater underestimation. This systematic deviation indicates that the inherent bias toward lower values observed for the SE~\cite{Evans2013} was not entirely eliminated by the modifications incorporated in the SEGWE model, as we further demonstrate in the Supporting Information.

For the mixture diolane~-~hexadecane, ESE does not adequately predict the temperature dependence of $D^{\infty}_{ij}$, which is underestimated. However, ESE still predicts the steepest slope with respect to temperature among the studied models, thereby most closely reflecting the experimental trend. The overestimation of the experimental data point at the lowest temperature may be due to its proximity to the melting point of hexadecane, which may affect diffusion in a way the model does not account for.

\FloatBarrier

The presented results demonstrate the predictive power of ESE and its applicability for reliably estimating diffusion coefficients at infinite dilution as a function of temperature across a wide range of binary mixtures. Although these findings were obtained for mixtures with previously unseen solutes, the model is also well-suited for predicting $D^{\infty}_{ij}$ in unseen solvents and in mixtures with unseen solutes \textit{and} solvents, as we demonstrate in the Supporting Information. The predictions spanning a wide temperature range, as shown in Figure~\ref{fig_Exp}, confirm the physical consistency of the results obtained with ESE.

\section{Conclusions}

Information on diffusion coefficients in liquid mixtures is crucial for modeling mass transfer and simulating separation processes. However, even for the simplest type, namely diffusion coefficients of single solutes infinitely diluted in pure solvents $D^{\infty}_{ij}$, experimental data are extremely scarce. Moreover, all previously existing models for predicting $D^{\infty}_{ij}$, including physical, semi-empirical, and machine learning (ML) approaches, are generally limited to specific solvent classes or mixtures containing only components with available experimental data, exhibit poor prediction accuracy, or fail to deliver physically consistent predictions.

In the present work, we address this challenge by introducing the Enhanced Stokes\discretionary{-}{}{}Einstein~(ESE) model. This hybrid approach combines the physical Stokes-Einstein~(SE) model with a neural network that learns systematic corrections to the deficiencies of the SE formulation. In this way, ESE enables reliable predictions of $D^{\infty}_{ij}$, while retaining the physical interpretability of the original SE model. In addition to the inputs required by the SE model, which are often readily available, ESE requires only readily available structural information about the solute and solvent molecules, thereby enabling broad applicability. The predictive performance of ESE was demonstrated using unseen test data for solutes and solvents that were deliberately withheld during model development and training. In all cases, ESE clearly outperforms existing prediction methods for $D^{\infty}_{ij}$, including the SEGWE model~\cite{Evans2018}.

The current version of ESE was trained on data for organic molecules and water, with constituent atoms no heavier than chlorine and molar masses below 1000~g~mol$^{-1}$. Although predictions outside this domain are possible, their use is discouraged. However, these limitations are not intrinsic to the model itself but reflect the availability of suitable experimental data. Extending the applicability of ESE requires substantially larger, more diverse, and higher-quality datasets than are currently available. In addition, ESE does currently not cover the diffusion of ionic species. Extending the model in this direction would be appealing, but would require reconsidering the underlying physical model.

Beyond the direct prediction of $D^{\infty}_{ij}$, the versatility of the ESE framework opens up further opportunities. In particular, it could be applied to inverse prediction problems, in which properties of unknown solutes, such as their molar mass $M_i$, are inferred from diffusion data. Such an approach would be especially attractive for the development of rational modeling strategies for poorly specified mixtures in conjunction with nuclear magnetic resonance~(NMR) fingerprinting techniques~\cite{Specht2021, Specht2023a, Specht2023b, Specht2023c, Wagner2025}.
\newpage

\section*{Conflicts of Interest}
There are no conflicts of interest to declare.

\section*{Data Availability}
The final trained ESE model is available on Zenodo at \url{https://doi.org/10.5281/zenodo.18787099}.

\begin{acknowledgement}

We gratefully acknowledge financial support by the Carl Zeiss Foundation in the projects 'Process Engineering 4.0' and 'Halocycles', as well as by DFG in the frame of the Research Training Group 'WERA' (project number 503479768), the Core Facility 'LASE-MR' (project number 537627671) and the Emmy Noether Group of FJ (project number 528649696). Model training was carried out on the high performance computer Elwetrisch at RPTU under the grant RPTU-MLVT.

\end{acknowledgement}

\begin{suppinfo}

Generation of molecular descriptors; experimental data measured in this work; hyperparameter study; comparison with MCM and TCM; supplementary results; results for solvent-split cross-validation

\end{suppinfo}

\bibliography{achemso-demo}

\end{document}

\section{Generation of Molecular Descriptors}

Molecular descriptors for solutes and solvents were automatically generated from their SMILES strings using RDkit~\cite{RDKit2024}. Table~\ref{tab_RDk} lists the RDKit functions used to obtain the information on the molecules for the generation of molecular descriptors. The numbers of non-hydrogen atoms and halogen atoms were determined from the molecule's atom list obtained using \texttt{rdkit.Chem.rdchem.Atom.GetAtoms}.

\begin{table}
\centering
\caption{RDkit~\cite{RDKit2024} functions used to obtain the molecular information to generate the molecular descriptors.}
\label{tab_RDk}
    \begin{tabular}{ll}
        \toprule
        Function & Molecular information\\
        \midrule
        \texttt{rdkit.Chem.Descriptors.ExactMolWt} & Molar mass  \\
        \texttt{rdkit.Chem.Lipinski.RingCount} & Number of rings in molecule \\
        \texttt{rdkit.Chem.Lipinski.NumHeteroatoms} & Number of heteroatoms in molecule \\
        \texttt{rdkit.Chem.Lipinski.NumHAcceptors} & Number of hydrogen-bond acceptors in molecule \\
        \texttt{rdkit.Chem.Lipinski.NumHDonors} & Number of hydrogen-bond donors in molecule \\
        \bottomrule
    \end{tabular}
\end{table}
\FloatBarrier

For water, the functions from Table~\ref{tab_RDk} erroneously yield values of zero for both hydrogen-bond acceptors and donors. To correct this, we manually set $r_\mathrm{Acc} = 0.500$ and $r_\mathrm{Don} = 0.500$ for water.

\section{Experimental Data Measured in This Work}

Diffusion coefficients at infinite dilution $D^{\infty}_{ij}$ were measured by $^{1}$H~PFG~NMR experiments at $T = 298.15$~K with mixtures containing low solute concentrations and extrapolation to the state of infinite dilution of the solutes. The experimental procedure was exactly the same as described in detail in our prior work~\cite{Wagner2025}. Details on the chemicals used for these experiments are given in Table~\ref{tab_chem}. Deionized and purified water was prepared using an ultrapure water system (Omnia series, stakpure). The numerical data for the new experimental $D^{\infty}_{ij}$ are provided in Table~\ref{tab_D}.

\begin{table}
\centering
\caption{Suppliers and purities (as specified by the suppliers) of the used chemicals.}
\label{tab_chem}
    \begin{tabular}{lllc}
        \toprule
        Chemical & Formula & Supplier & Purity \\
         &  &  & \% \\
        \midrule
        Acetonitrile & C$_2$H$_3$N & Sigma-Aldrich & $\geq$ 99.90 \\
        Decane & C$_{10}$H$_{22}$ & TCI & $\geq$ 99.00 \\
        Diglyme & C$_6$H$_{14}$O$_3$ & TCI & $\geq$ 99.00 \\
        Dimethyl sulfoxide & C$_2$H$_6$OS & Merck & $\geq$ 99.90 \\
        2,6-Dimethylpyridine & C$_7$H$_9$N & Sigma-Aldrich & $\geq$ 99.90 \\
        L(+)-Ascorbic acid & C$_6$H$_8$O$_6$ & Roth & $\geq$ 99.00 \\
        Mandelic acid & C$_8$H$_8$O$_3$ & Sigma-Aldrich & $\geq$ 99.90 \\
        \bottomrule
    \end{tabular}
\end{table}
\FloatBarrier

\begin{table}
\centering
\caption{Diffusion coefficients at infinite dilution $D^{\infty}_{ij}$ at $T = 298.15$~K measured using PFG NMR spectroscopy.$^\mathrm{a}$}
\label{tab_D}

\begin{threeparttable}
\begin{tabular}{llc}
\toprule
Solute~$i$ & Solvent~$j$ & \multicolumn{1}{c}{$D^{\infty}_{ij}$} \\
 & & 10$^{9}$ m$^{2}$ s$^{-1}$ \\
\midrule
    2,6-Dimethylpyridine & water & 0.71 \\
    L(+)-Ascorbic acid & water & 1.13 \\
    Mandelic acid & water & 0.76 \\
    Acetonitrile & decane & 3.93 \\
    Acetonitrile & diglyme & 1.87 \\
    Dimethyl sulfoxide & decane & 1.92 \\
    Dimethyl sulfoxide & diglyme & 1.45 \\
\bottomrule
\end{tabular}

\begin{tablenotes}
\footnotesize
\item[a] Relative uncertainty of diffusion coefficient
\mbox{$\Delta D^{\infty}_{ij} = 0.02$~$\cdot$ 10$^{9}$ m$^{2}$ s$^{-1}$}.
Uncertainty of temperature
\mbox{$\Delta T = 0.1$~K}.
\end{tablenotes}
\end{threeparttable}
\end{table}
\FloatBarrier

\section{Hyperparameter Study}

Table~\ref{tab_hyper} gives an overview of the hyperparameters of ESE varied in the present work, i.e., the AdamW optimizer’s weight decay $\lambda$, the initial learning rate, batch size, and the number of layers and nodes in the NN, along with the performance of each variant in terms of the mean validation loss (MSRE) across all folds. Model 1, which achieved the lowest validation loss, was identified as the best hyperparameter configuration and used throughout the manuscript. However, the results in Table~\ref{tab_hyper} demonstrate that the performance of ESE is robust with respect to variations in the hyperparameters. 

\begin{table}
\centering
\caption{Tested ESE variants with varied hyperparameters and respective mean validation loss (MSRE). $\lambda$ is the weight decay of the AdamW optimizer.}
\label{tab_hyper}
\begin{adjustbox}{width=\textwidth}
    \begin{tabular}{lllllll}
        \toprule
        Model No. & $\lambda$ & Initial learning rate & Batch size & Layers & Nodes & Mean validation loss\\
        \midrule
        1  & $1 \cdot 10^{-3}$ & 0.001 & 4  & 2 & 32 and 16 &  0.029 \\
        2  & $1 \cdot 10^{-4}$ & 0.001 & 4  & 2 & 32 and 16 &  0.030 \\
        3  & $1 \cdot 10^{-2}$ & 0.001 & 4  & 2 & 32 and 16 &  0.030 \\
        4  & $1 \cdot 10^{-3}$ & 0.0001 & 4  & 2 & 32 and 16 &  0.029 \\
        5  & $1 \cdot 10^{-3}$ & 0.01 & 4  & 2 & 32 and 16 &  0.029 \\
        6  & $1 \cdot 10^{-3}$ & 0.001 & 2  & 2 & 32 and 16 &  0.030 \\
        7  & $1 \cdot 10^{-3}$ & 0.001 & 8  & 2 & 32 and 16 &  0.031 \\
        8  & $1 \cdot 10^{-3}$ & 0.001 & 4  & 1 & 32 and 16 &  0.032 \\
        9  & $1 \cdot 10^{-3}$ & 0.001 & 4  & 3 & 32 and 16 &  0.031 \\
        10  & $1 \cdot 10^{-3}$ & 0.001 & 4  & 2 & 24 and 12 &  0.032 \\
        11  & $1 \cdot 10^{-3}$ & 0.001 & 4  & 2 & 40 and 20 &  0.031 \\
        \bottomrule
    \end{tabular}
\end{adjustbox}
\end{table}
\FloatBarrier

\section{Comparison of ESE with MCM and TCM}

To ensure a fair comparison of ESE with the MCM~\cite{Romero2025} and TCM~\cite{Romero2026} methods, we retrained and reevaluated the MCM and TCM models using the subset of $N = 391$ data points from our data set that fall within their applicability domain at 298.15~K, 313.15~K, and 333.15~K, adopting the same hyperparameter settings and leave-one-out testing procedure described in the original publications. For the MCM, independent models were trained at each temperature, and their results were combined to evaluate the method.
For ESE, we used the same hyperparameter settings and procedures described in the manuscript. However, we evaluated the model by leaving out all data associated with a single system, rather than all data associated with a single solute, to match the evaluation protocol used for the MCM and TCM.

\begin{figure}
\centering
    \includegraphics[width=8.0cm]{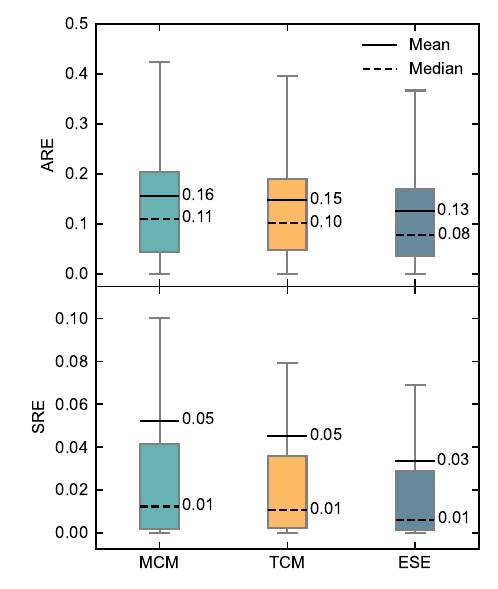}
    \caption{Boxplots of the ARE and SRE of the predicted diffusion coefficients at infinite dilution from MCM~\cite{Romero2025}, TCM~\cite{Romero2026}, and ESE, evaluated on the subset of data within the scope of the MCM and TCM. The box width indicates the interquartile range, and the whisker length is 1.5 times the interquartile range. Outliers are not depicted for visual clarity.}
    \label{fig_MCMTCM}
\end{figure}

Figure~\ref{fig_MCMTCM} shows the ARE and SRE of the predictions from MCM, TCM, and ESE, demonstrating the superior predictive performance of ESE over the other models.
Note that the $N = 391$ data points considered in this comparison represent only about 39\% of the available $D^\infty_{ij}$ data within the substantially broader applicability domain of ESE.

\FloatBarrier

\section{Supplementary Results}

\begin{figure}
\centering
    \includegraphics[width=12.5cm]{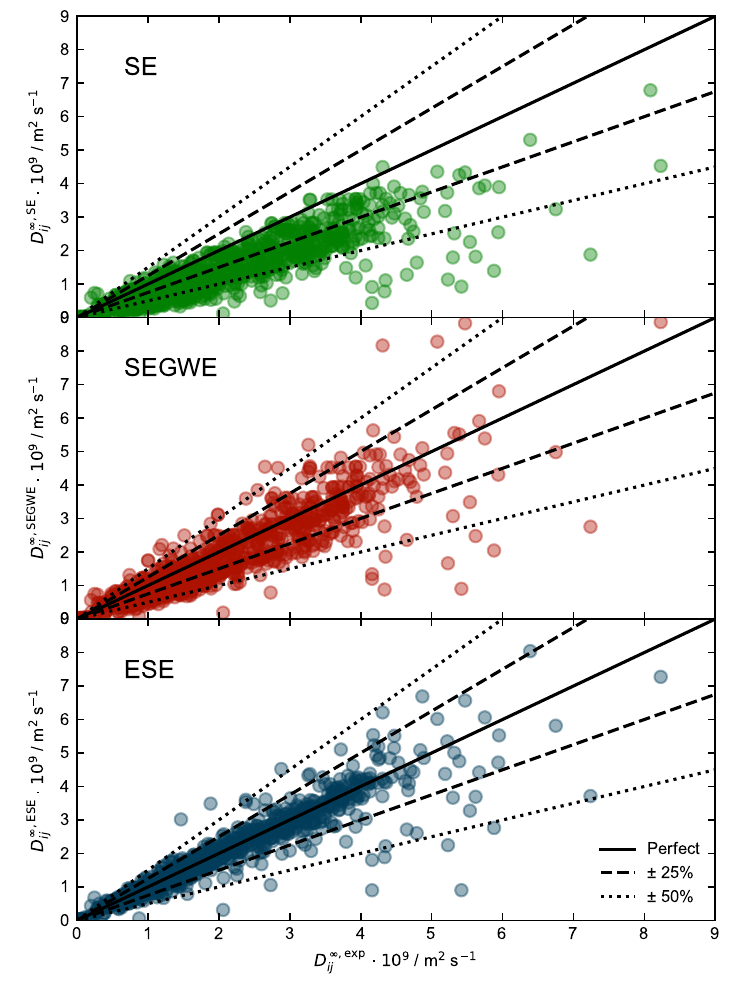}
    \caption{Comparison of the predicted (pred) diffusion coefficients at infinite dilution~$D^{\infty}_{ij}$ obtained with SE, SEGWE, and ESE to the corresponding experimental (exp) test data.}
    \label{fig_Par}
\end{figure}

Figure~\ref{fig_Par} shows parity plots of the predicted $D^{\infty}_{ij}$ from the test data over the corresponding experimental values for SE, SEGWE, and ESE. In the case of the SE, the predictions~$D^{\infty,\mathrm{SE}}_{ij}$ systematically underestimate the experimental values, which aligns with the findings of Evans et al.~\cite{Evans2013}.
For SEGWE, the extent of this underprediction is reduced, although a slight bias remains, particularly at higher experimental values~$D^{\infty,\mathrm{exp}}_{ij}$. In contrast, ESE does not show a systematic deviation from the experimental values and generally has smaller prediction errors.

\FloatBarrier

\section{Results for Solvent-Split Cross-Validation}

\begin{figure}
\centering
    \includegraphics[width=8.0cm]{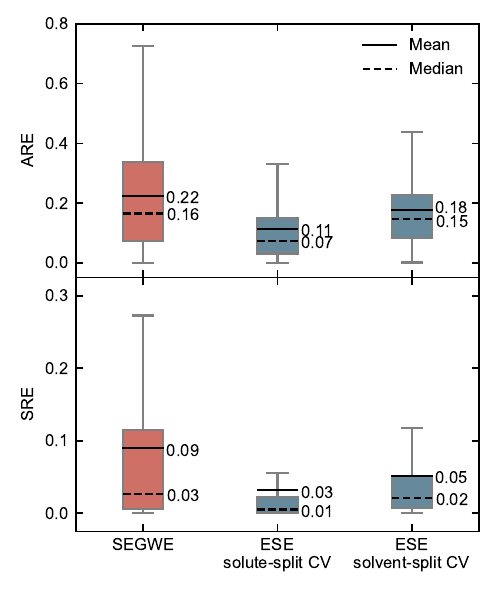}
    \caption{Boxplots of the ARE and SRE of the predicted diffusion coefficients at infinite dilution from SEGWE and ESE. For ESE, the results obtained on unseen solutes (solute-split CV, cf. manuscript) are compared to those obtained on unseen solvents (solvent-split CV). The box width indicates the interquartile range, and the whisker length is 1.5 times the interquartile range. Outliers are not depicted for visual clarity.}
    \label{fig_Sol}
\end{figure}

Figure~\ref{fig_Sol} compares the ARE and SRE of the results obtained when using a solvent-split cross-validation~(CV) during training and evaluation of ESE to the results shown in the manuscript, for which a solute-split CV was employed.
The predictive performance of ESE decreases when tested on mixtures with unseen solvents instead of unseen solutes.
The results indicate a more challenging task of predicting~$D^{\infty}_{ij}$ for mixtures with unknown solvents, which could be explained by the fact that our data set contains fewer unique solvents compared to solutes (42 instead of 209), making it more difficult for the model to learn the influence of different solvents on~$D^{\infty}_{ij}$.
However, even in this case, ESE clearly outperforms the SEGWE model, demonstrating its suitability for predicting~$D^ {\infty}_{ij}$ in mixtures with unseen components.

\FloatBarrier

\newpage

\bibliography{achemso-demo}